\documentstyle[aps,prl,graphicx]{revtex}

\draft

\begin{document}

\title{Electric-octupole and pure-electric-quadrupole effects in soft-x-ray photoemission}

\author{A.\ Derevianko,$^{1,}$\cite{an} O.\ Hemmers,$^2$ S.\ Oblad,$^2$
P.\ Glans,$^3$ H.\ Wang,$^4$ S.\ B.\ Whitfield,$^5$ R.\ Wehlitz,$^6$
I.\ A.\ Sellin,$^7$ W.\ R.\ Johnson,$^1$ and D.\ W.\ Lindle$^2$}

\address{$^1$Department of Physics, University of Notre Dame, Notre Dame, IN 46556}
\address{$^2$Department of Chemistry, University of Nevada, Las Vegas, NV 89154-4003}
\address{$^3$Atomic Physics, Stockholm University, 10405 Stockholm, Sweden}
\address{$^4$Department of Physics, Uppsala University, Box 530, S-751 21 Uppsala, Sweden}
\address{$^5$Department of Physics, University of Wisconsin, Eau Claire, WI 54702}
\address{$^6$Synchrotron Radiation Center, University of Wisconsin, Stoughton, WI 53589}
\address{$^7$Department of Physics, University of Tennessee, Knoxville, TN 37996}

\date{\today}

\maketitle

\begin{abstract}
Second-order [$O(k^2)$, $k=\omega/c$] nondipole effects in soft-x-ray photoemission are
demonstrated via an experimental and theoretical study of angular distributions of neon valence
photoelectrons in the 100--1200~eV photon-energy range.  A newly derived theoretical
expression for nondipolar angular distributions characterizes the second-order effects using four
new parameters with primary contributions from pure-quadrupole and octupole-dipole
interference terms.  Independent-particle calculations of these parameters account for a
significant portion of the existing discrepancy between experiment and theory for Ne $2p$
first-order nondipole parameters.    
\end{abstract}

\pacs{PACS: 32.80.Fb, 31.25.Eb}

\twocolumn

A mainstay of photoemission is the (electric-)dipole approximation (DA), in which all
higher-order multipoles are neglected \cite{JES1}.  The range of validity of the DA has received
renewed interest as recent experiments \cite{Krassig,Hemmers} uncovered breakdowns at
progressively lower photon energies.  At high energies ($\hbar\omega>$5 keV), breakdown of
the DA in photoionization is well-known, and a proper description requires inclusion of many
multipoles \cite{JES9}.  For soft-x-ray ($\hbar\omega<$5 keV) photoionization, in contrast,
first-order [$O(k)$] corrections to the DA generally have been considered sufficient
\cite{JESpaper}.  At these relatively low energies, DA breakdown primarily leads to
forward/backward asymmetries in photoelectron angular-distribution patterns.  Especially
striking have been observations of nondipole effects at energies below 1 keV \cite{Hemmers}, a
region in which the DA is usually considered valid.  In the present work, an experimental and
theoretical analysis of neon valence photoemission demonstrates a new, and unexpected,
breakdown: significant second-order [$O(k^2)$] nondipole effects, primarily due to electric-octupole and pure-electric-quadrupole interactions, in low-energy photoemission.

We begin with Cooper's $O(k)$ formula for the differential photoionization cross section of a
subshell $(n,\kappa)$ in a randomly oriented target using linearly polarized light \cite{coop}:
\begin{eqnarray}
\lefteqn{\frac{d\sigma_{n\kappa}}{d\Omega} = \frac{\sigma_{n\kappa}}{4\pi} 
\Bigl\{\, 1 + \beta_{n\kappa}\, P_2(\cos{\theta}) \Bigr. }\hspace{1in} \nonumber \\ 
 && \Bigl. +\, \left( \delta_{n\kappa} + \gamma_{n\kappa} \, \cos^2{\theta} \right) \sin{\theta} 
\, \cos{\phi} \Bigr\}  , \label{dxs}
\end{eqnarray} 
where  $\sigma_{n\kappa}$ is the photoionization cross section, $\beta_{n\kappa}$ describes
the angular distribution within the DA, and $\delta_{n\kappa}$ and $\gamma_{n\kappa}$ are
nondipole angular-distribution parameters characterizing the leading first-order corrections to the
DA (mostly $E_2-E_1$ terms).  The angles $\theta$ and $\phi$ are determined by the direction
of the photoelectron relative to the photon-polarization {\boldmath $\hat{\epsilon}$} and
photon-propagation {\boldmath $k$} directions, respectively.  The first two terms on the right of 
Eq.~(\ref{dxs}) constitute the usual DA expression for the differential cross section, and the DA
notion of a "magic angle" [$\theta_m= 54.7^\circ$, $P_2(\cos{\theta_m}) = 0$] is preserved only in
the $\phi=90^\circ$ plane perpendicular to {\boldmath $k$}.

At this level of approximation, recent rare-gas experiments \cite{Krassig,Hemmers} observed
significant modifications of photoelectron angular distributions from DA expectations, generally
in good agreement with first-order independent-particle-approximation (IPA) calculations
\cite{coop,BP}.  The only exception is Ne $2p$ \cite{Hemmers}; while measured values of
$\gamma_{2s}$ ($\delta_{2s}$ is negligible when $\beta_{2s}=2$) agree fairly well with
calculations, measured values of the combined parameter $\zeta_{2p}$
($=3\delta_{2p} + \gamma_{2p}$) are 30\% larger than IPA predictions for energies near 1 keV.

The same experiment also found $\beta_{2p}$ disagrees substantially with IPA calculations in
this energy region, but is in close agreement with correlated calculations using the random-phase
approximation (RPA) \cite{JL,Amusia}, thereby identifying important electron-correlation
effects well above the $n=2$ thresholds \cite{JES49}.  This result led to speculation
\cite{Hemmers} the discrepancy between measured and IPA-calculated $\zeta_{2p}$ values
might also be due to interchannel-coupling effects.  However, subsequent first-order nondipole
calculations including electron correlation \cite{JDCDM} disproved this notion; RPA values of
Ne $\zeta_{2p}$ are in excellent agreement with the uncorrelated IPA results \cite{coop,BP}.

In this work, we explain much of this discrepancy between theory and experiment for Ne
$\zeta_{2p}$.  Beginning with theory, second-order [$O(k^2)$] corrections to the differential
cross section, which arise from interferences between $E_1-E_3$,  $E_1-M_2$, $E_2-E_2$,
$E_2-M_1$, and $M_1-M_1$,  and from retardation corrections to $E_1-E_1$ amplitudes, are
incorporated into Eq.~(\ref{dxs}):
\begin{eqnarray}
\lefteqn{ \frac{d\sigma_{n\kappa}}{d\Omega}  = \frac{\sigma_{n\kappa}}{4\pi} \Bigl\{ 1 +
\left(\beta_{n\kappa} + \Delta\beta_{n\kappa} \right)\, 
P_2(\cos{\theta}) \Bigr. }\hspace{0.0in} \nonumber \\
&&  +\, \left( \delta_{n\kappa} +\gamma_{n\kappa} \cos^2{\theta} \right)\,  
\sin{\theta}\cos{\phi} +\, \eta_{n\kappa} \, P_2(\cos{\theta})\cos{2\phi} \nonumber \\[0.1pc]
&& \hspace{0.4in} \Bigl. + \, \mu_{n\kappa}\, \cos{2\phi}
+ \xi_{n\kappa}\, \left( 1+\cos{2\phi} \right)\, P_4(\cos{\theta}) \Bigr\} \, , \label{dxs1}
\end{eqnarray}
where the $O(k^2)$-parameters $\Delta\beta , \eta, \mu$, and $\xi$ are introduced
\cite{AtDataOk2}.  Three of them satisfy the constraint $\eta+\mu+\xi=0$. 
Reference~\cite{AtDataOk2} contains complete formulae and a tabulation of first- and
second-order parameters for all subshells of the rare gases helium to xenon. 

For experiment, we first present results for Ne $\gamma_{2s}$ and $\zeta_{2p}$ determined
assuming only first-order corrections, embodied in $\delta$ and $\gamma$, are needed to
correctly interpret the data.  The experiments were performed with an apparatus designed to
measure deviations from the DA \cite{RSI}; $\gamma_{2s}$ and $\zeta_{2p}$ are determined
using angle-resolved photoemission intensities at $\theta_m$ and both $\phi=0^\circ$ and
$\phi=90^\circ$.  Figure~1 compiles old \cite{Hemmers} and new values for $\zeta_{2p}$ and
$\gamma_{2s}$ (open squares) determined in this way.  The solid curves represent $O(k)$
calculations \cite{coop,BP,JDCDM}, which agree well with the $2s$ results, but disagree with
the $2p$ results above 800 eV.

To obtain $O(k^2)$ predictions (dotted curves in Fig.~1), we first carried out numerical IPA
studies of second-order corrections for neon.  Wavefunctions for bound-state and continuum
electrons were obtained from the radial Dirac equation in a modified Hartree potential.  Values
for all $n=2$ angular-distribution parameters were calculated up to 2 keV.  Our results for
$\gamma_{2s}$ and $\zeta_{2p}$ are in excellent agreement with previous nonrelativistic IPA
calculations \cite{coop}.  Figure~2 shows values for $\Delta\beta$, $\eta$, and $\mu$
($\xi=-\eta-\mu$) obtained from our second-order IPA calculations.  Primary contributions to
these parameters come from $E_1-E_3$ and $E_2-E_2$ terms, with the octupole term
contributing about 65\% at 1 keV.  A smaller contribution ($\approx 10 \%$) comes from the
$E_1-M_2$ term.

For our measurement geometry \cite{RSI}, Eq.~(\ref{dxs1}) and our $O(k^2)$ calculations can
be used to estimate the influences of second-order effects on the analysis of experimental results
for $\zeta_{2p}$ ($=3\delta_{2p} + \gamma_{2p}$).  Specifically, an effective value of $\zeta$,
including these influences, can be defined as 
\begin{eqnarray}
(\zeta)_{\rm eff} &=& \sqrt{\frac{27}{2}}
\left[ \frac{\sigma( \theta_m,0 ) }
{\sigma (\theta_m,\pi/2) } -1 \right]  \nonumber \\[0.5pc] 
&\approx& 
\frac{ \gamma+3\delta +\sqrt{54} (  \mu - 7 \xi/18)}{1-\mu} \, , \label{two} 
\end{eqnarray}
using the notation $\sigma(\theta,\phi) = \frac{d\sigma}{d\Omega}(\theta,\phi)$.  Using the
results in Fig.~2, $\zeta_{\rm eff}$ for $2p$ and $\gamma_{\rm eff}$ for $2s$ can be
determined and compared to the results in Fig.~1 to identify influences of second-order
nondipole effects.  (We stress the measurements are fine, only the assumptions behind their
analysis are suspect.)  This exercise yields the dotted curves in Fig.~1, providing excellent
agreement with $\gamma_{2s}$ and clearly improved agreement with $\zeta_{2p}$.  The
second-order corrections contained in $\zeta_{\rm eff}$ account for much of the difference
between the first-order theory and experiment for $\zeta_{2p}$, demonstrating the first
observation of $O(k^2)$ effects in soft-x-ray photoemission.  

To confirm this unexpected finding, we made additional measurements with our apparatus,
which contains four electron analyzers in a chamber which can rotate about the photon beam.  At
a nominal angular position, two analyzers are at $\theta_m$ and $\theta=0^\circ$ in the plane
perpendicular to the photon beam ($\phi=90^\circ$), which we refer to as the dipole plane because
first-order corrections vanish, while the other analyzers are positioned on the $35.3^\circ$ cone in the
forward direction with respect to the photon beam, with one of them at $\theta_m$ and
$\phi=0^\circ$.  Photoemission intensities in the magic-angle analyzers are independent of $\beta$
and can differ only because of nondipole effects.  While the magic angle is no longer strictly
valid when second-order effects are included, calculations show they can be unimportant in
certain geometries (see below).  

New measurements were performed at ten rotational positions, yielding 20 angle-resolved
intensities for Ne $2s$ and $2p$ photoemission at different angles $\theta$ within the dipole
plane, and 20 more at different angles $\theta$ and $\phi$ around the nondipole cone.  From the
calculated results for $\Delta \beta_{2p}$ in Fig.~2, direct second-order effects on $\beta_{2p}$
should be insignificant; $\Delta \beta_{2p}\approx0.005$ near 1 keV, much smaller than our
measurement uncertainties.  Therefore, values of $\beta_{2p}$ determined from the dipole-plane
spectra should agree well with DA calculations, {\em if} effects due to $\eta$, $\mu$, and $\xi$
are negligible in the dipole plane.  Using Fig.~2, the influence of these parameters on angle-resolved photoemission intensities can be predicted.  In the dipole plane, we predict their effects
will mostly cancel, and thus the excellent agreement \cite{JES49} between experiment and
theory for $\beta_{2p}$ is not surprising.

In the nondipole cone, influences of the second-order parameters are superimposed on intensity
variations due to the dipole $\beta$ and the first-order $\delta$ and $\gamma$ parameters. 
However, for both $2s$ and $2p$, our calculations predict effects due to $\eta$, $\mu$, and $\xi$
also mostly cancel around the nondipole cone.  Furthermore, small residual effects around this
cone are similar in sign and magnitude for $2s$ and $2p$, which is relevant because intensity
ratios are the raw input for data analysis.  Assuming no influence of second-order effects in the
nondipole cone, we modeled the measured $2s/2p$ ratios around this cone using Eq.~(\ref{dxs})
to derive values for $\gamma_{2s}$ and $\zeta_{2p}$.  These results (solid circles in Fig.~1)
agree extremely well with first-order calculations \cite{coop,BP,JDCDM}, confirming our
prediction of near cancellation of second-order effects in this geometry.

The above confirmation tests include two independent experimental methods to determine
$\gamma_{2s}$ and $\zeta_{2p}$: one relies on measurements at many angles in the nondipole
cone, the other relies on comparison of magic-angle-only measurements in the dipole plane and
the nondipole cone.  For the former, second-order effects mostly vanish.  For the latter, in
contrast, the influences of $\eta$, $\mu$, and $\xi$ on Ne $2p$ photoemission are expected to be
opposite in sign for $\phi=0^\circ$ and $\phi=90^\circ$, because of the $\cos(2\phi)$ terms in
Eq.~(\ref{dxs1}).  Thus, second-order effects should be observable with the latter method, hence
the differences in values for $\zeta_{2p}$ derived using the two methods.

As a demonstration of the influence of second-order nondipole effects on angle-resolved-photoemission intensities, Fig.~3 compares spectra taken with the two magic-angle analyzers. 
Figure 3a contains a neon photoemission spectrum taken at $\hbar\omega$=1200 eV,
$\theta_m$, and $\phi=90^\circ$ in the dipole plane, where influences of $\beta$, $\delta$, and
$\gamma$ vanish.  Included are fit curves showing modeled peak shapes and photoemission
satellites to the left of the $2s$ peak.  The overall fit (solid curve) matches the data very well, as
indicated by the residual in Fig.~3b.

This spectrum and fit are reproduced in Fig.~3c and compared to a nondipole-cone spectrum at
1200 eV, $\theta_m$, and $\phi=0^\circ$.  Intensity normalization between the spectra was achieved
using $\gamma_{2s}$ from Fig.~1, for which experiment and theory agree well.  By inspection,
the $2s/2p$ ratio is different in the two spectra.  One possible explanation is a differential
influence of first-order nondipole effects on the $2s$ and $2p$ intensities.  As a quantitative test
of this hypothesis, we derived the dotted region in Fig.~3c by multiplying the fit to the $2p$ peak
in the dipole-plane spectrum by the expected differential effect on $2s$ and $2p$ peak intensities
in the nondipole-cone spectrum determined from IPA-/RPA-predicted values for
$\gamma_{2s}$ and $\zeta_{2p}$ (see Eq.~(\ref{dxs})).  If first-order effects alone explain the
observed variation of the $2s/2p$ ratio, then the dotted region should coincide with the $2p$
peak in the nondipole-cone spectrum.  It does not, and thus the difference between the dotted
region and the open-circle data ($\approx 10\%$) is attributed to the influence of second-order
effects.

In conclusion, an experimental and theoretical study of valence photoemission from neon has
demonstrated the first observation of second-order (primarily $E_1-E_3$ and $E_2-E_2$)
nondipole effects on photoelectron angular distributions in the soft-x-ray region.  A general
expression for the differential photoionization cross section, including all contributions through
second order, has been derived in a form convenient for comparison to experiment.

This work was supported by NSF and DOE EPSCoR.  AD and WRJ were supported in part by
NSF grant PHY-99-70666.  DWL acknowledges UNLV Sabbatical Leave support.  The
experiments were performed at the Advanced Light Source at Lawrence Berkeley National Lab.,
supported by the DOE Materials Science Division, BES, OER under contract
DE-AC03-76SF00098.

\begin{figure}
\centerline{\includegraphics[scale=0.6]{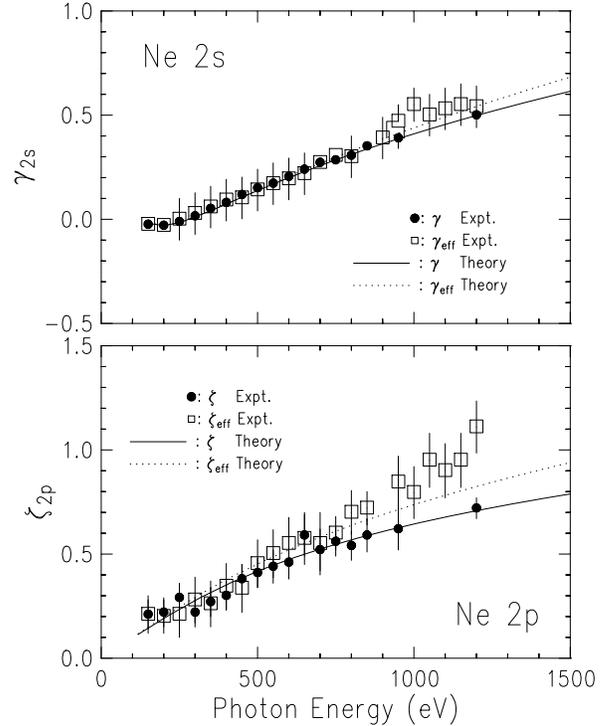}}
\caption{Experimental and theoretical values of $\gamma_{2s}$ and $\zeta_{2p}$
($3\delta_{2p} + \gamma_{2p}$) for neon. The effective quantities include second-order
[$O(k^2)$] nondipole influences.  See text for complete description.
\label{fig1}}
\end{figure}

\begin{figure}
\centerline{\includegraphics[scale=0.6]{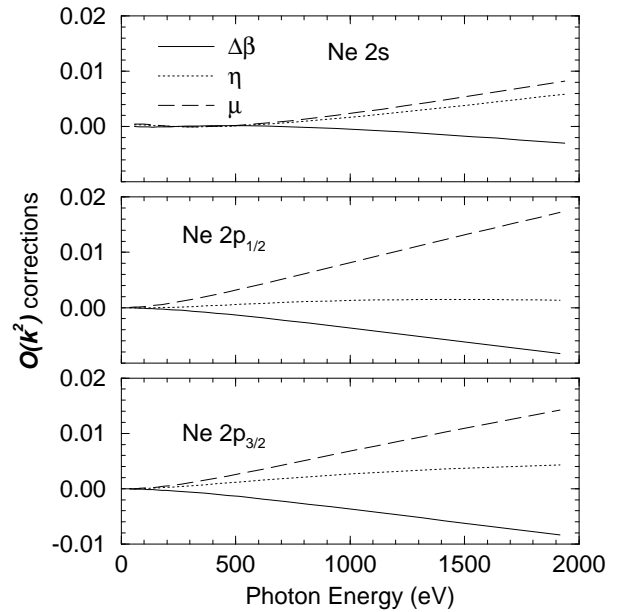}}
\caption{Nondipole parameters of $O(k^2)$ for neon.
\label{fig2}}
\end{figure}

\begin{figure}
\centerline{\includegraphics[scale=0.45]{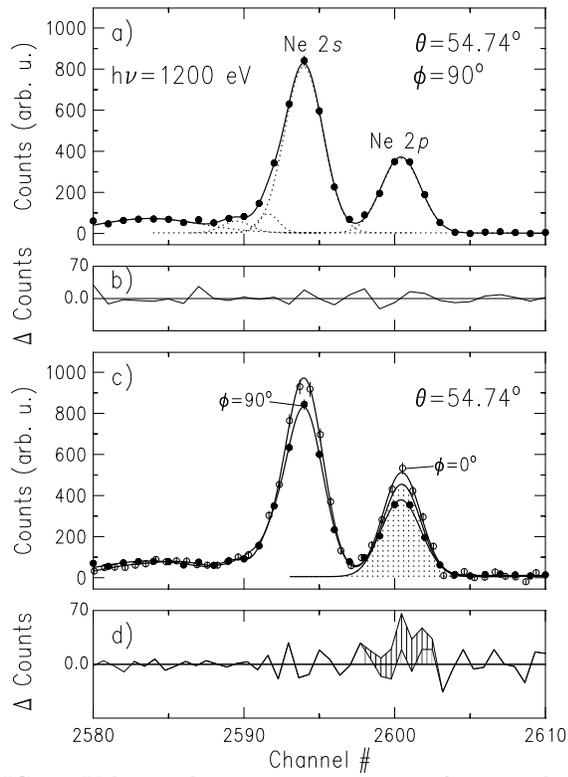}}
\caption{Valence photoemission spectra of neon taken at 1200~eV and $\theta_m= 54.7^\circ$.  a)
$\phi=90^\circ$ spectrum, including a fit.  b) Residual of fit in (a).  c) Same spectrum and fit as in
(a) compared to a $\phi=0^\circ$ spectrum.  d) Residual of fit to the $\phi=0^\circ$ spectrum (lower
curve), and difference between the $\phi=0^\circ$ fit and the dotted region in (c) (upper curve).  The
hatched area is $2p$ photoemission intensity attributable to second-order corrections.  See text for
full explanation.
\label{fig3}}
\end{figure}
\end{document}